\documentclass[12pt]{article}
\usepackage[a4paper=true]{hyperref}
\usepackage{amsfonts}
\usepackage{amssymb}
\usepackage{amsmath}
\usepackage{amsthm}
\usepackage[table]{xcolor}
\usepackage{color}

\hypersetup{
    colorlinks = true,
    linkcolor = red,
    anchorcolor = red,
    citecolor = blue,
    filecolor = red,
    urlcolor = red
}
\pagecolor{white}

\voffset = - 1.2cm
\begin{document}

\centerline{\bf \Large On Lie systems and Kummer--Schwarz equations} 
\vskip 0.25cm
\centerline{J. de Lucas$^\dagger$ and C. Sard\'on$^\ddagger$}
\vskip .5cm
\centerline{Faculty of Mathematics and Natural Sciences, Cardinal Stefan Wyszsy\'nski University,}
\centerline{ W\'oycickiego 1/3, 01-938, Warsaw, Poland}%
\centerline{Department of Fundamental Physics, Faculty of Sciences, University of 
Salamanca,}\centerline{Plza. de la Merced s/n, 37.008, Salamanca, Spain
}%

\begin{abstract}

A {\it Lie system} is a system of first-order differential equations admitting a
{\it superposition rule}, i.e., a map that expresses its
general solution in terms of a generic family of particular solutions and
certain constants. In this work, we use the geometric 
theory of Lie systems to prove that the explicit integration of second- and third-order Kummer--Schwarz
equations is equivalent to
obtaining a particular solution of a Lie system on
$SL(2,\mathbb{R})$. This same result can be extended to Riccati, Milne--Pinney and other related equations. We demonstrate that all the above-mentioned equations associated with exactly the same Lie system on $SL(2,\mathbb{R})$ can be integrated simultaneously. This retrieves and generalizes in a unified and simpler manner previous results appearing in the literature. 
As a byproduct, we recover various properties of the {\it Schwarzian derivative}.
\end{abstract}


\section{Introduction}

Geometric techniques have been proved to be a very successful approach to
solve
differential equations, leading to methods of key importance in physics and
mathematics, such as: Lie symmetries, the Painlev\'e method, and Lax pairs
\cite{Lax,Lax2,Painleve1,Painleve2}. 

In this work, we focus on the geometric theory of Lie systems
\cite{LS,Ve93,Ve93II,Ve94,Gu93,PW,CGM00,CGM07}.
Lie systems have lately attracted some attention owing to their
numerous applications and properties \cite{Dissertationes}. For instance, they have been employed to study the integrability of Riccati and matrix Riccati equations \cite{CarRam,HWA83,LW96}, control
 and Floquet theory \cite{EV11,FE08,FLV10,Ru10}, and other equations appearing in classical and quantum mechanics
\cite{Dissertationes}. Furthermore, their generalizations have led to
the geometric investigation of stochastic equations \cite{JP09}, superequations
\cite{BecGagHusWin90}, and other
topics \cite{CarRam03,CGL11}. 

More specifically, we analyze second- and
third-order Kummer--Schwarz equations \cite{GGG11,AL08,Be82,Be88} -- henceforth KS-2 and KS-3 equations
 -- with the aid of the theory of Lie systems. The mathematical relevance
of these equations resides in their close connection with the 
Kummer's
problem \cite{Be88,BR97,Be07}, the study of homogeneous systems of second-order differential equations \cite{Cr10},
the Schwarzian derivative \cite{OT09}, and other related themes \cite{Be82,Be88,BR97,Ta89,As10,Ma94,Ta97}.
Moreover, some
interest has been focused on the study of solutions of KS-2 and KS-3 equations, which have been analyzed in several manners in the
literature: e.g., through
non-local transformations or in terms of solutions to other differential
equations \cite{GGG11,AL08,Be82,BR97,Be07}. From a physical viewpoint, KS-2 and
KS-3  
equations occur in the study of Milne--Pinney equations, Riccati
equations, and time-dependent frequency harmonic oscillators \cite{CGL11,GGG11,AL08,Co94}, which are of certain relevancy in
two-body problems \cite{Be81,Be89}, quantum mechanics \cite{IK03,Kr09}, 
classical mechanics \cite{NR02}, etcetera \cite{Pe05}. 

First, we show that KS-2 and KS-3 equations can be
studied through two $\mathfrak{sl}(2,\mathbb{R})$--Lie systems \cite{Pi12}, i.e., Lie systems that describe the integral curves of a
$t$-dependent vector field taking values in a Lie algebra of vector fields -a so-called {\it Vessiot--Guldberg Lie algebra}- isomorphic to $\mathfrak{sl}(2,\mathbb{R})$. This new result slightly generalizes previous findings about these equations \cite{CGL11}. 
 
Afterwards, we obtain two Lie group actions whose
fundamental vector fields correspond with those of the above-mentioned Vessiot--Guldberg Lie algebras. These actions 
allow us to prove that the explicit integration of KS-2 and KS-3
equations is equivalent to working out a particular solution of a Lie system on
$SL(2,\mathbb{R})$. Further, we  will see that 
Riccati and Milne--Pinney equations exhibit similar features. 

We show that
the knowledge of the general solution of any of the abovementioned equations allows us to solve simultaneously any other related to the same equation on $SL(2,\mathbb{R})$. This fact provides a new powerful and general way of linking solutions of these equations, which were previously known to be related through {\it ad hoc} expressions in certain cases \cite{AL08,Co94}. Additionally, our approach can potentially be extended to other 
$\mathfrak{sl}(2,\mathbb{R})$-Lie systems \cite{Dissertationes}. 

Subsequently, we derive a superposition rule for certain Lie systems associated with a relevant family of KS-3 equations of the form $\{x,t\}=2b_1(t)$, with $b_1(t)$ being an arbitrary function of $t$ and $\{x,t\}$ the Schwarzian derivative of a function $x(t)$ with respect to $t$ \cite{Be82,OT09}. This permits us
to recover some features of these remarkable equations \cite{OT09} and the Schwarzian derivatives.
In particular, we prove that the general solution of $\{x,t\}=2b_1(t)$ can be described by means of an expression that depends on a particular solution and several constants related to the initial conditions, i.e., a {\it basic superposition rule for higher-order differential equations} \cite{CGL11}. 
This enables us to solve some relevant cases of these equations. 
Additionally, our basic superposition rule allows us 
to retrieve some known symmetries of Schwarzian derivatives \cite{OT09}. 

The direct Lie--Scheffers theorem \cite{GL12} states that every Lie system
possesses a {\it mixed superposition rule}, i.e., a type of functions
that allows us to express its general solution in terms of particular solutions
of (possibly different) systems of first-order differential equations
and some constants \cite{GL12}. We here derive a mixed superposition rule to investigate general
solutions of KS-2 equations in terms of $t$-dependent
frequency harmonic oscillators. This mixed superposition rule generalizes a previous result described by Berkovich \cite{Be07}. It is also 
remarkable that mixed superposition rules can be applied to describe
general solutions of second- and third-order Kummer--Schwarz equations in terms of other 
$\mathfrak{sl}(2,\mathbb{R})$-Lie systems.
An example is to reproduce expressions describing the general solutions of $\{x,t\}=2b_1(t)$ in terms of particular solutions of Riccati equations or certain time-dependent frequency harmonic oscillators \cite{Ne49}. 

The structure of the paper goes as follows. Section II is devoted to surveying
several concepts used throughout the paper. In Section III and IV, we address the
analysis of KS-2 and KS-3
equations
and show how their integration can be reduced to solving certain Lie systems on
$SL(2,\mathbb{R})$. Section
V
is
dedicated to the analysis of connections of second- and third-order Kummer--Schwarz equations with other equations. In
Section VI we accomplish a new approach to the Schwarzian derivative based upon our results. Subsequently, we provide a new superposition rule for KS-3 equations  in Section VII. To conclude, we resume our results and describe the future work to do in Section VIII. 

\section{Fundamentals}\label{Fun}
For simplicity, we hereafter assume all mathematical objects, e.g., vector
fields 
and superposition rules, to be real, smooth, and globally defined on vector
spaces. In this manner,
we
highlight the key points of our presentation by omitting the analysis of minor
technical problems (see \cite{CGM00,CGM07,CL10SecOrd2} for details).

\subsection{On $t$-dependent vector fields}
The geometrical study of Lie systems is based upon the notion of {\it
$t$-dependent vector fields} \cite{FM}. Given the tangent bundle projection
$\tau:{\rm T}\mathbb{R}^n\rightarrow\mathbb{R}^n$ and the projection
$\pi_2:(t,x)\in\mathbb{R}\times\mathbb{R}^n\mapsto x\in\mathbb{R}^n$, a {\it
$t$-dependent vector field} $X$ on $\mathbb{R}^n$ is a map
$X:(t,x)\in\mathbb{R}\times\mathbb{R}^n\mapsto X(t,x)\in \mathbb{\rm
T}\mathbb{R}^n$ satisfying that $\tau\circ X=\pi_2$. This condition implies that
 $X$ can be considered as a family
$\{X_t\}_{t\in\mathbb{R}}$ of vector fields $X_t:x\in\mathbb{R}^n\mapsto
X_t(x)=X(t,x)\in{\rm T}_x\mathbb{R}^n$ and vice versa \cite{Dissertationes}.  

As standard vector fields, $t$-dependent vector fields also admit integral
curves. We
call {\it integral curve} of $X$ an integral curve
$\gamma:\mathbb{R}\rightarrow \mathbb{R}\times \mathbb{R}^n$ 
of the {\it suspension} of $X$, i.e., the vector field on $\mathbb{R}\times N$
given by $\partial/\partial t+X(t,x)$  \cite{FM}.

From a modern geometric viewpoint, every system of first-order differential
equations 
\begin{equation}\label{LieSystem}
\frac{dx^i}{dt}=X^i(t,x),\qquad i=1,\ldots,n,
\end{equation}
can be associated with the unique $t$-dependent vector field on $\mathbb{R}^n$,
namely, 
\begin{equation}\label{TVF}
X(t,x)=\sum_{i=1}^nX^i(t,x)\frac{\partial}{\partial x^i},
\end{equation}
whose integral curves are (up to an appropriate reparametrization) of the form
 $(t,x(t))$, with $x(t)$ being a solution of system (\ref{LieSystem}). Conversely, 
 every $t$-dependent vector field (\ref{TVF}) determines a unique
system of first-order differential equations, the so-called {\it associated
system}, determining its integral curves of the form $(t,x(t))$. This justifies
to denote by $X$ both a $t$-dependent vector field and its associated
system.
\subsection{Lie systems and superposition rules}

\label{Def1b} A {\it superposition rule}  for a system $X$ on
$\mathbb{R}^n$ is a map $\Phi:(\mathbb{R}^{n})^m\times\mathbb{R}^{n}\rightarrow
\mathbb{R}^n$ of the form
$$u=\Phi({u_{(1)}},\ldots,{u_{(m)}};k_1,\ldots,k_{n}),$$
such that the general solution $x(t)$ of $X$ can be written as
\begin{equation}\label{SupHODE}
x(t)=\Phi\left({x_{(1)}}(t),\ldots,{x_{(m)}}(t);k_1,\ldots,k_{n}\right),
\end{equation}
with $x_{(1)}(t),\ldots, x_{(m)}(t)$ being a generic family of particular
solutions and $k_1,\ldots,k_{n}$ being a set of constants related to the initial
conditions of the system. 

The characterization of systems of first-order differential equations admitting
a superposition rule was obtained by Lie \cite{LS}. Its result, the nowadays
called {\it Lie--Scheffers Theorem}, is the 
cornerstone of the theory of Lie systems \cite{CGM07,BM09II} and related theories \cite{Dissertationes,GL12}.

The Lie--Scheffers theorem states that a system $X$ possesses a
superposition rule if and only if 
\begin{equation}\label{LieDecom}
X_t=\sum_{\alpha=1}^rb_\alpha (t)Y_\alpha,
\end{equation}
for a certain family $Y_1,\ldots,Y_r$ of vector fields spanning an
$r$-dimensional real Lie algebra, the so-called associated {\it Vessiot--Guldberg
Lie algebra},
and $t$-dependent functions $b_1(t),\ldots,b_r(t)$.

Every Lie system $X$ associated with a Vessiot--Guldberg Lie algebra $V$  gives
rise
to a (generally local) Lie group action $\varphi:G\times
\mathbb{R}^n\rightarrow\mathbb{R}^n$ whose fundamental vector fields are the elements of $V$ and such that $T_eG\simeq V$ with $e$ being the neutral element of $G$ \cite{Palais}. This action
allows us to write the general solution of $X$, which can be assumed to be of the form (\ref{LieDecom}), as
\begin{equation}\label{mix}
x(t)=\varphi(g_1(t),x_0), \qquad x_0\in\mathbb{R}^n,
\end{equation}
with $g_1(t)$ being a particular solution of
\begin{equation}\label{eqLie}
\frac{dg}{dt}=-\sum_{\alpha=1}^rb_\alpha(t)Y_\alpha^R(g),
\end{equation}
where $Y^R_1,\ldots,Y_r^R$ is a certain basis of right-invariant vector fields on $G$
such that $Y^R_\alpha(e)={\rm a_\alpha}\in T_eG$, with $\alpha=1,\ldots,r$, and each ${\rm a_\alpha}$ is the element of $T_eG$ associated to the fundamental vector field $Y_\alpha$ (see \cite{CGM00} for
details). 

Since $Y^R_1,\ldots,Y_r^R$ span a finite-dimensional real Lie algebra, the Lie--Scheffers Theorem guarantees that (\ref{eqLie}) admits a superposition rule and becomes a Lie system. Indeed, as the right-hand
side of (\ref{eqLie}) is invariant under right-translations, its general solution
can be brought into the form
\begin{equation}\label{SupGroup}
g(t)=R_{g_0}g_1(t),\qquad g_0\in G,
\end{equation}
where $g_1(t)$ is a particular solution of (\ref{eqLie}) and $R_{g_0}$, with $g_0\in G$, is the map $R_{g_0}:g'\in G\mapsto g'\cdot g_0\in G$ \cite{CGM00}. 
In other words, (\ref{eqLie}) admits a superposition rule.

Finally, given the general solution of $X$, the solution $g_1(t)$ of the associated (\ref{eqLie}) with $g_1(0)=e$ can
be characterized as the unique solution to the algebraic system $x_p(t)=\varphi(g_1(t),x_p(0))$, where $x_p(t)$ ranges over a ``sufficient large set'' of
particular solutions of $X$ \cite{PW}.
\subsection{Mixed superposition rules}

Despite its theoretical relevance, expression (\ref{mix}) only becomes useful to obtain
explicitly the  
 general solution of $X$ in terms of a particular solution of
(\ref{eqLie}) or vice versa, provided the explicit form of $\varphi$ is known. Unfortunately, this
is usually a complicated task as, for instance, in the case of most autonomous Lie
systems \cite{CL10SecOrd2}. Nevertheless, expression (\ref{mix}) is also interesting for
constituting
a particular example of a {\it mixed superposition rule}, which generalizes the notion of superposition rules \cite{GL12}. 

A {\it mixed
superposition rule} for a system $X$ is a $(m+1)$-tuple
$(\Phi,X_{(1)},\ldots,X_{(m)})$ consisting of a function
$\Phi:\mathbb{R}^{n_1}\times\ldots\times\mathbb{R}^{n_m}\times
\mathbb{R}^{n}\rightarrow\mathbb{R}^{n}$ and a series of systems $X_{(a)}$ on
$\mathbb{R}^{n_a}$, with $a=1,\ldots,m$, such that the general solution $x(t)$
of $X$ can be cast in the form
\begin{equation}\label{MixedSup}
x(t)=\Phi(x_{(1)}(t),\ldots,x_{(m)}(t);k_1,\ldots,k_{n}),
\end{equation}
where $x_{(1)}(t),\ldots,x_{(m)}(t)$ is a generic family of particular solutions
of $X_{(1)},\ldots,X_{(m)}$, respectively, 
and $k_1,\ldots,k_{n}$ are real constants.

The {\it direct Lie--Scheffers Theorem} \cite{GL12} states that a system $X$ admits
a mixed superposition rule if and only if $X$ is a Lie system. Although this restricts the
use of mixed superposition rules to Lie systems, the notion is still relevant, as mixed superposition rules are in general more versatile
and easier to derive than standard ones \cite{GL12}.
  
Let us now describe a procedure to derive mixed superposition rules. This
method is based on the so-called direct product of $t$-dependent vector fields and the
geometrical characterization of mixed superposition rules as foliations (we refer to \cite{Dissertationes,GL12,SIGMA} for details and examples).

\label{Ext} Given a certain family $X_{(0)},\ldots,\!X_{(m)}$ of
$t$-dependent vector fields defined, respectively, on
$\mathbb{R}^{n_0},\ldots,\mathbb{R}^{n_m}$, their {\it direct product} (or {\it
direct prolongation}) $Z\equiv X_{(0)}\times\cdots\times X_{(m)}$ is the unique
$t$-dependent vector field $Z$ on
$\mathbb{R}^{n_0}\times\ldots\times\mathbb{R}^{n_m}$ such that ${\rm
pr}_{a*}Z_t=(X_{(a)})_t$, with ${\rm
pr}_a:(x_{(0)},\ldots,x_{(m)})\in\mathbb{R}^{n_0}\times\ldots\times\mathbb{R}^{
n_m}\mapsto x_{(a)}\in \mathbb{R}^{n_a}$ for $a=0,\ldots,m$ and all
$t\in\mathbb{R}$. 

It can be proved that every mixed superposition rule $(\Phi,X_{(1)}, \ldots,  X_{(m)})$ for a system $X$ on $\mathbb{R}^n$
gives rise to an $n$-codimensional foliation $\mathcal{F}$ on
$\mathbb{R}^{n_1}\times\ldots\times\mathbb{R}^{n_m}\times\mathbb{R}^n$ whose  
leaves $\mathcal{F}_k$, with $k=(k_1,\ldots,k_n)\in\mathbb{R}^n$, are of the form  
$$
\mathcal{F}_k=\{(x_{(1)},\ldots,x_{(m)},x)|x=\Phi(x_{(1)},\ldots,x_{(m)};k)\}.
$$
Such leaves project diffeomorphically onto
$\mathbb{R}^{n_1}\times\ldots\times\mathbb{R}^{n_m}$ via
\begin{equation}\label{proj}
\begin{array}{lccc}
{\rm pr}:&\mathbb{R}^{n_1}\times\ldots\times
\mathbb{R}^{n_m}\times\mathbb{R}^n&\longrightarrow &
\mathbb{R}^{n_1}\times\ldots\times \mathbb{R}^{n_m}\\
&(x_{(1)},\ldots,x_{(m)},x)&\mapsto &(x_{(1)},\ldots,x_{(m)})
\end{array}
\end{equation}
and the vector fields $\{(X_{(1)}\times\ldots\times X_{(m)}\times X)_t\}_{t\in\mathbb{R}}$ are
tangent to them. Conversely, it is known that a foliation of the above type
also gives rise to a mixed superposition rule $(\Phi,X_{(1)},\ldots,X_{(m)})$ for $X$ \cite{GL12}.

We turn to describing a procedure to construct a foliation
of the above type and, from it, a mixed superposition rule for a system $X$. In
view of the direct Lie--Scheffers theorem,
this is only possible if $X$ is a Lie system. If so, assume $V$ 
to be an associated Vessiot--Guldberg Lie algebra. Take a
basis $Y_1,\ldots,Y_r$ of $V$. Determine a family $V_{(1)},\ldots,V_{(m)}$ of
Lie algebras of vector fields on
$\mathbb{R}^{n_1},\ldots,\mathbb{R}^{n_m}$, respectively, isomorphic to $V$ and
admitting a series of bases 
$$
Y^{(a)}_1,\ldots,Y^{(a)}_r\in V^{(a)},\qquad a=1,\ldots,m,
$$
that satisfy the same commutating relations as $Y_1,\ldots,Y_r$ and such that
the vector fields $
\widehat
Y_\alpha\equiv Y^{(1)}_\alpha\times\ldots \times Y^{(m)}_\alpha$, with
$\alpha=1,\ldots,r,
$ are linearly independent at a generic point (note that we have $r\leq \sum_{a=1}^mn_{a}$). 
It is  
important to remark that
these bases can easily be found in the literature of Lie systems
\cite{Dissertationes},
which facilitates the derivation of mixed superposition rules. 
Indeed, it
will be posteriorly shown that bases of this type 
appear naturally throughout this work.

From the above bases, construct a set of
vector fields 
$
\widetilde Y_\alpha=\widehat Y_\alpha\times Y_\alpha$ on
$\mathbb{R}^{n_1}\times\ldots\times
\mathbb{R}^{n_m}\times\mathbb{R}^n$, with $\alpha=1,\ldots,r
$. These vector fields span a real Lie algebra
$\widetilde V$  of dimension
$r\leq \sum_{a=1}^mn_a$. Furthermore, their elements span an integrable
distribution $\mathcal{D}$ on $\mathbb{R}^{n_1}\times\ldots\times
\mathbb{R}^{n_m}\times\mathbb{R}^n$ of rank $r$ of the form
$$
\mathcal{D}_\xi=\{Z(\xi)\mid Z\in \widetilde V\}\subset {\rm T}_\xi
(\mathbb{R}^{n_1}\times\ldots\times \mathbb{R}^{n_m}\times\mathbb{R}^n),
$$
where $Z(\xi)$ is the value of the vector field $Z$ at $\xi$.
Besides, $\mathcal{D}$ projects under (\ref{proj}) onto an integrable
distribution on
$\mathbb{R}^{n_1}\times\ldots\times \mathbb{R}^{n_m}$ 
spanned by the vector fields $\widehat Y_1,\ldots,\widehat Y_r$ and ${\rm
pr}_{*\xi}:\mathcal{D}_\xi\subset {\rm T}_\xi(\mathbb{R}^{n_1}\times\ldots\times
\mathbb{R}^{n_m}\times\mathbb{R}^n)\rightarrow {\rm T}_{{\rm
pr}(\xi)}(\mathbb{R}^{n_1}\times\ldots\times \mathbb{R}^{n_m})$ is an injective
mapping for a generic $\xi\in \mathbb{R}^{n_1}\times\ldots\times
\mathbb{R}^{n_m}\times\mathbb{R}^n$. 

As $\mathcal{D}$ is an integrable distribution of rank $r\leq \sum_{a=1}^m n_a$ 
 defined on a manifold of dimension $\sum_{a=1}^m n_a+n$, it can be proved \cite{Dissertationes,GL12} that
there exist $n$
functionally independent first-integrals
$F_1,\ldots,F_n$ for all vector fields of $\mathcal{D}$ satisfying that
\begin{equation}\label{cond}
\frac{\partial (F_1,\ldots,F_n)}{\partial (x^1,\ldots,x^n)}\neq 0,\qquad x\equiv
(x^1,\ldots,x^n)\in\mathbb{R}^n,
\end{equation}
at a generic point of $\mathbb{R}^{n_1}\times\ldots\times\mathbb{R}^{n_m}\times\mathbb{R}^n$.
Hence, given a point $(x_{(1)},\ldots,x_{(m)})$ and certain constants
$(k_1,\ldots,k_n)$, there exists a unique $x\in\mathbb{R}^n$ such that
$F_i(x_{(1)},\ldots,x_{(m)},x)=k_i$, for $i=1,\ldots,n$.  
For every $k=(k_1,\ldots,k_n)$, the solutions of the equations $F_i=k_i$, with $i=1,\ldots,n$, define
the points of a leaf $\mathcal{F}_k$, of an $n$-codimensional foliation $\mathcal{F}$ whose leaves project diffeomorphically onto
$\mathbb{R}^{n_1}\times\ldots\times \mathbb{R}^{n_m}$.
Further, the vector fields $Z_t=\sum_{\alpha=1}^rb_\alpha(t)\widetilde
Y_\alpha$, are tangent
to the leaves of $\mathcal{F}$. 

Note that $Z=X_{(1)}\times\ldots X_{(m)}\times X$, with $X=\sum_{\alpha=1}^rb_\alpha(t)Y_\alpha$  and $X_{(a)}=\sum_{\alpha=1}^rb_\alpha(t)Y^{(a)}_\alpha$ for $a=1,\ldots,m$.
In view of the characterization of mixed superposition rules as projectable foliations and using that the vector fields $\{Z_t\}_{t\in\mathbb{R}}$ are tangent to the leaves of $\mathcal{F}$, we obtain
that $\mathcal{F}$ describes a mixed 
superposition rule for $X$   depending on
particular 
solutions of the systems $X_{(a)}$,
with $a=1,\ldots,m$. In fact, by considering the equations $F_i=k_i$, with $i=1,\ldots,n$ and
$k_1,\ldots,k_n$ being certain real constants, we can describe the
value of $x=(x^1,\ldots,x^n)$ in terms of $k_1,\ldots,k_n$ and $x_{(1)},\ldots,x_{(m)}$. It can be proved that the resulting expressions determine a map
\begin{equation}\label{SupMapping}
\begin{array}{lccc}
\Phi:&\mathbb{R}^{n_1}\times\ldots\times
\mathbb{R}^{n_m}\times\mathbb{R}^n&\longrightarrow &
\mathbb{R}^{n}\\
&(x_{(1)},\ldots,x_{(m)};k_1,\ldots,k_n)&\mapsto&(x^1,\ldots,x^n)
\end{array}
\end{equation}
that permits us to express the general solution $x(t)$ of $X$ in the form (\ref{MixedSup})
in terms of a generic
family of particular solutions of the systems $X_{(1)},\ldots,X_{(m)}$ and real
constants
$k_1,\ldots,k_n$, i.e., $(\Phi,X_{(1)},\ldots,X_{(m)})$ is a mixed superposition
rule for $X$. 

Observe that if we additionally impose that 
$V_{(1)}=\ldots=V_{(m)}=V$ and choose the same basis $Y_1,\ldots,Y_r$ for
all these Lie algebras, we recover the procedure for describing standard
superposition rules
detailed in \cite{CGM07}. Indeed, in this case, we obtain a family of vector fields $\widehat
Y_\alpha=Y_\alpha\times\ldots\times Y_\alpha (m-{\rm times})$, with
$\alpha=1,\ldots,r$, the so-called {\it diagonal prolongations}
\cite{CGM07,GL12} of $Y_\alpha$ to
$\mathbb{R}^{mn}$. When $m$ is such that the vector fields
$\widehat Y_1,\ldots,\widehat Y_r$ are linearly independent (at a generic point), 
the determination
of $n$ common functionally independent first-integrals for $\widetilde
Y_\alpha=\widehat Y_\alpha\times Y_\alpha$, with $\alpha=1,\ldots,r,$ satisfying
(\ref{cond}) gives rise to a superposition rule for $X$ depending on $m$
particular solutions.

\section{The second-order Kummer--Schwarz equations}
Let us now turn to analyzing KS-2 equations. These
equations take the form
\begin{equation}\label{KS2}
\frac{d^2x}{dt^2}=\frac 3{2x}\left(\frac{dx}{dt}\right)^2-2c_0x^3+2b_1(t)x,
\end{equation}
with $c_0$ being a real constant and $b_1(t)$ a $t$-dependent function. 
KS-2
equations are a particular case of second-order Gambier equations
\cite{CGL11,GGG11} and appear in the study of cosmological models \cite{NR02}. 
In addition, their relations to other differential equations like Milne--Pinney
equations \cite{GGG11}, make them an
alternative approach to the analysis of many physical problems
\cite{CGL11,AL08,IK03}. 

Consider the first-order system 
\begin{equation}\label{FirstOrderKummer}
\left\{\begin{aligned}
\frac{dx}{dt}&=v,\\
\frac{dv}{dt}&=\frac 32 \frac{v^2}x-2c_0x^3+2b_1(t)x,
\end{aligned}\right.
\end{equation}
on ${\rm T}\mathbb{R}_0$, with
$\mathbb{R}_0=\mathbb{R}-\{0\}$, obtained by adding the new variable
$v\equiv dx/dt$ to the KS-2 equation (\ref{KS2}). This
system describes the integral curves of the $t$-dependent vector field
\begin{equation}\label{Dec0}
\!X_t=v\frac{\partial }{\partial x}+\left(\frac 32
\frac{v^2}x\!-\!2c_0x^3+2b_1(t)x\right)\frac{\partial }{\partial
v}\!=\!M_3+b_1(t)M_1,
\end{equation}
where 
\begin{equation}\label{VFKS2}
\begin{gathered}
M_1=2x\frac{\partial}{\partial v},\qquad M_2=x\frac{\partial}{\partial
x}+2v\frac{\partial }{\partial v},\qquad M_3=v\frac{\partial}{\partial
x}+\left(\frac 32\frac{v^2}x-2c_0x^3\right)\frac{\partial}{\partial v}
\end{gathered}
\end{equation}
satisfy the commutation relations
\begin{equation}
[M_1,M_3]=2M_2,\quad [M_1,M_2]=M_1,\quad [M_2,M_3]=M_3.
\end{equation}
These vector fields span a three-dimensional real Lie algebra $V$ of vector fields
isomorphic to $\mathfrak{sl}(2,\mathbb{R})$\cite{Dissertationes,CGL11}.  Hence, in view of (\ref{Dec0}) and the 
Lie--Scheffers Theorem, $X$ admits a superposition rule and becomes a Lie system
associated with a Vessiot--Guldberg Lie algebra isomorphic to $\mathfrak{sl}(2,\mathbb{R})$, i.e.,
a $\mathfrak{sl}(2,\mathbb{R})$-Lie system. 

As shown in Section \ref{Fun}, the knowledge of 
a Lie group action $\varphi_{2KS}:G\times {\rm T}\mathbb{R}_0\rightarrow {\rm T}\mathbb{R}_0$ whose fundamental vector fields are $V$ and $T_eG\simeq V$ allows us to express
 the general solution of $X$ in
the form (\ref{mix}), in terms of a particular solution of a Lie system (\ref{eqLie}) on
$G$. Let us determine $\varphi_{2KS}$ in such a
way that our procedure can easily be extended to third-order
Kummer--Schwarz equations. 

Consider the basis of matrices of $\mathfrak{sl}(2,\mathbb{R})$ 
\begin{equation}\label{Base}
{\rm a}_1=\left(
\begin{array}{cc}
0&1\\
0&0 
\end{array}
\right),\quad {\rm a}_2=\frac 12\left(
\begin{array}{cc}
-1&0\\
0&1
\end{array}
\right)
,\quad {\rm a}_3=\left(
\begin{array}{cc}
0&0\\
-1&0 
\end{array}
\right)
\end{equation}
satisfying the commutation relations
$$
[{\rm a}_1,{\rm a}_3]=2{\rm a}_2,\qquad [{\rm a}_1,{\rm a}_2]={\rm a}_1,\qquad
[{\rm a}_2,{\rm a}_3]={\rm a}_3,
$$
which match those satisfied by $M_1,M_2$ and $M_3$. So, the linear
function $\rho:\mathfrak{sl}(2,\mathbb{R})\rightarrow V$ mapping ${\rm
a}_\alpha$ into
$M_\alpha$, with $\alpha=1,2,3$, is a Lie algebra isomorphism. If we consider it as
an infinitesimal Lie group action, we can then ensure that there
exists a local Lie group action $\varphi_{2KS}:SL(2,\mathbb{R})\times {\rm
T}\mathbb{R}_0\rightarrow{\rm T}\mathbb{R}_0$ obeying the required properties. In particular, 
$$
\frac{d}{ds}\varphi_{2KS}(\exp(-s{\rm a}_\alpha),{\bf
t}_{x})=M_\alpha(\varphi_{2KS}(\exp(-s{\rm a}_\alpha),{\bf t}_{x})), 
$$
where ${\bf t}_{x}\equiv (x,v)\in {\rm T}_{x}\mathbb{R}_0\subset {\rm
T}\mathbb{R}_0$, $\alpha=1,2,3$, and $s\in\mathbb{R}$. This condition determines the action
on ${\rm T}\mathbb{R}_0$ of the
elements of $SL(2,\mathbb{R})$ of the form
$\exp(-s{\rm a}_\alpha)$, with $\alpha=1,2,3$ and $s\in\mathbb{R}$.  By integrating $M_1$ and $M_2$, we obtain
\begin{equation}\label{FunEl}
\begin{aligned}
\varphi(\exp_{2KS}(-\lambda_1{\rm a}_1),{\bf
t}_{x})&=\left(x,v+2x\lambda_1\right),\\
\varphi(\exp_{2KS}(-\lambda_2{\rm a}_2),{\bf
t}_{x})&=\left(xe^{\lambda_2},ve^{2\lambda_2}
\right).\\
\end{aligned}
\end{equation}
Observe that $M_3$ is not defined on ${\rm T}_0\mathbb{R}$. So, its integral curves, let us
say $(x(\lambda_3),v(\lambda_3))$, must be fully contained in either ${\rm
T}\mathbb{R}^+$ or ${\rm T}\mathbb{R}^-$. These integral curves are
determined by the system
\begin{equation}\label{LC}
\frac{dx}{d\lambda_3}=v,\qquad \frac{dv}{d\lambda_3}=\frac 32\frac {v^2}x-2c_0x^3.
\end{equation}
When $v\neq 0$, we obtain
$$
\frac{dv^2}{dx}=\frac{3v^2}{x}-4c_0x^3\Longrightarrow
v^2(\lambda_3)=x^3(\lambda_3) \Gamma-4c_0x^4(\lambda_3),
$$
for a real constant $\Gamma$. Hence, for each integral curve $(x(\lambda_3),v(\lambda_3))$, we have 
$$
\Gamma=\frac{v^2(\lambda_3)+4c_0x^4(\lambda_3)}{x^3(\lambda_3)}.
$$
Moreover, it easy to see that $d\Gamma/d\lambda_3=0$ not only for solutions of (\ref{LC}) with $v(\lambda_3)\neq 0$ for every $\lambda_3$, but for any solution of (\ref{LC}). 
Using the above results and (\ref{LC}), we
see that
\begin{equation}\label{cur}
\frac{dx}{d\lambda_3}={\rm sg}(v)\sqrt{\Gamma x^3-4c_0x^4 }\Rightarrow
x(\lambda_3)=
\frac{x(0)}{F_{\lambda_3}(x(0),v(0))},
\end{equation}
where ${\rm sg}$ is the well-known {\it sign function} and
$$
F_{\lambda_3}({\bf t}_{x})=\left(1-\frac{v\lambda_3}{2x}\right)^2+
c_0x^2\lambda_3^2.
$$
Now, from (\ref{cur}) and taking into account the first equation within (\ref{LC}),
it immediately follows that 
\begin{equation}\label{FunEl2}
\begin{aligned}
\varphi_{2KS}(\exp(-\lambda_3{\rm a}_3),{\bf
t}_{x})&=\left(\frac{x}{F_{\lambda_3}({\bf
t}_{x})},\frac{v-\frac{
v^2+4c_0x^4}{2x}\lambda_3}{F_{\lambda_3}^{2}({\bf t}_{x})}\right)\!.
\end{aligned}
\end{equation}

Let us employ previous results to determine the action  on $\mathfrak{sl}(2,\mathbb{R})$
of those
elements $g$ close to the neutral element $e\in SL(2,\mathbb{R})$. 
Using the so-called {\it canonical coordinates of the second kind} \cite{It87},
we can write $g$ within an open neighborhood $U$ of $e$ in a unique form as
\begin{equation}\label{decomposition}
g=\exp(-\lambda_3{\rm a}_3)\exp(-\lambda_2{\rm a}_2)\exp(-\lambda_1{\rm a}_1),
\end{equation}
for real constants $\lambda_1,\lambda_2$ and $\lambda_3$. This allows us to
obtain the action of every
$g\in U$  on ${\rm T}\mathbb{R}_0$ through the composition of the actions of
elements $\exp(-\lambda_\alpha
{\rm a}_\alpha)$, with $\lambda_\alpha\in\mathbb{R}$ for $\alpha=1,2,3$. 
To do so, we determine the
constants
$\lambda_1,\lambda_2$ and $\lambda_3$ associated to each $g\in U$ in (\ref{decomposition}).

Considering the standard matrix representation of $SL(2,\mathbb{R})$, we can
express every
$g\in SL(2,\mathbb{R})$ as
\begin{equation}\label{rep}
g=\left(\begin{array}{cc}
\alpha &\beta\\
\gamma &\delta 
\end{array}\right),\qquad \alpha\delta-\beta\gamma=1,\qquad
\alpha,\beta,\gamma,\delta\in\mathbb{R}.
\end{equation}
In view of (\ref{Base}), and comparing (\ref{decomposition}) and (\ref{rep}), we obtain
$$
\alpha=e^{\lambda_2/2},\quad \beta=-e^{\lambda_2/2}\lambda_1,\qquad
\gamma=e^{\lambda_2/2}\lambda_3. 
$$
Consequently, 
$$
\lambda_1=-\beta/\alpha,\qquad \lambda_2=2\log \alpha,\qquad
\lambda_3=\gamma/\alpha, 
$$
and, from the basis (\ref{Base}), the decomposition (\ref{decomposition}) and expressions (\ref{FunEl}) and (\ref{FunEl2}), the action reads
$$
\varphi_{2KS}\left(g,
{\rm \bf t}_{x}\right)=\left(\frac{x}{F_g({\bf
t}_{x})},\frac{1}{F_g^{2}({\bf t}_{x})}\left[
(v\alpha-2x\beta)\left(\delta-\frac{\gamma v}{2x}\right)-2c_0x^3\alpha\gamma\right]\right),
$$
where 
$$
F_g({\bf
t}_{x})=\left(\delta-\frac{\gamma v}{2x}\right)^2
+c_0x^2\gamma^2.
$$
Although this expression has been derived for $g$ being close to $e$, it can be proved
that the action is properly defined at points $(g,{\bf t}_x)$ such that $F_g({\bf t}_x)\neq 0$. If $c_0>0$, then $F_{g}({\bf t}_x)>0$ for all $g\in SL(2,\mathbb{R})$ and ${\bf t}_x\in{\rm T}\mathbb{R}_0$. So, $\varphi_{2KS}$ becomes globally defined. Otherwise,
 $F_{g}({\bf t}_x)>0$ for $g$ close enough to $e$. Then, $\varphi_{2KS}$ is only defined on a neighborhood of $e$.

The action $\varphi_{2KS}$ also permits us to write the general solution of system
(\ref{FirstOrderKummer}) in the form $(x(t),v(t))=\varphi_{2KS}(g(t),{\bf
t}_{x})$,
with
 $g(t)$ being a particular solution of 
\begin{equation}\label{reduced}
\frac{dg}{dt}=-Y^R_3(g)-b_1(t)Y^R_1(g),
\end{equation}
where $Y^R_\alpha$, with $\alpha=1,2,3$, are the single right-invariant
vector fields on $SL(2,\mathbb{R})$ such that $Y_\alpha^R(e)={\rm a}_\alpha$ \cite{CGM00,Dissertationes}. 
Additionally, as $x(t)$ is the general solution of KS-2 equation
(\ref{KS2}), we readily see that
\begin{equation}\label{action}
x(t)=\tau\circ \varphi_{2KS}(g(t),{\bf t}_{x}),
\end{equation}
with $\tau:(x,v)\in{\rm
T}\mathbb{R}\mapsto x\in \mathbb{R}$ a tangent bundle projection, provides us with the general solution of (\ref{KS2})  in terms
of a particular
solution of (\ref{reduced}).

Conversely, we prove that
we can recover a particular solution to (\ref{reduced}) from the knowledge of the general solution of (\ref{KS2}). For simplicity, we will determine 
the particular solution $g_1(t)$ with $g_1(0)=e$. Given two particular solutions $x_1(t)$ and $x_2(t)$ of (\ref{KS2}) with $dx_1/dt(t)=dx_2/dt(t)=0$, the expression (\ref{action}) implies that
$$
(x_i(t),v_i(t))=\varphi_{2KS}(g_1(t),(x_i(0),0)),\qquad i=1,2.
$$
Writing the above expression explicitly, we get
\begin{equation}\label{sys1}
\begin{aligned}
-\frac{x_i(0)v_i(t)}{2x_i^2(t)}&=\beta(t)\delta(t)+c_0 x_i^2(0)\alpha(t)\gamma(t),\\
\frac{x_i(0)}{x_i(t)}&=\delta^2(t)+c_0x_i^2(0)\gamma^2(t),
\end{aligned}
\end{equation}
for $i=1,2$. 
The first two equations allow us to determine the value of $\beta(t)\delta(t)$ and $\alpha(t)\gamma(t)$. Meanwhile, we can obtain the value of $\delta^2(t)$ and $\gamma^2(t)$ from the other two ones. As $\delta(0)=1$, we know that $\delta(t)$ is positive when close to $t=0$. Taking into account that we have already worked out $\delta^2(t)$, we can determine $\delta(t)$ for small values of $t$. Since we have already obtained $\beta(t)\delta(t)$, we can also derive $\beta(t)$ for small values of $t$ by using $\delta(t)$. Note that $\alpha(0)=1$. So, $\alpha(t)$ is positive for small values of $t$, and the sign of $\alpha(t)\gamma(t)$ determines
the sign of $\gamma(t)$ around $t=0$. In view of this, the value of $\gamma(t)$ can be determined from $\gamma^2(t)$ in the interval around $t=0$. Summing up, we can obtain algebraically a particular solution of (\ref{sys1}) with $g_1(0)=e$ from the general solution of (\ref{KS2}). 

\section{The third-order Kummer--Schwarz equations} 
The results obtained in the previous section can be generalized directly to the
case of KS-3 equations, i.e., the third-order differential equations
\begin{equation}\label{KS3}
\frac{d^3x}{dt^3}=\frac 32\left(
\frac{dx}{dt}\right)^{-1}\!\!\left(\frac{d^2x}{dt^2}\right)^{2}\!\!-2c_0(x)\left(\frac{
dx}{dt}\right)^3\!\!+2b_1(t)\frac{dx}{dt},
\end{equation}
where $c_0=c_0(x)$ and $b_1=b_1(t)$ are arbitrary. 

The relevance of KS-3 equations resides in
their relation to
the Kummer's problem \cite{Be82,Be88,Be07}, Milne--Pinney \cite{AL08}
and Riccati
equations \cite{AL08,Co94,EEL07}. Such  relations can be useful in the
interpretation of physical systems through KS-3 equations, e.g., the case of
quantum
non-equilibrium dynamics of many body systems \cite{GBD10}. Furthermore, KS-3
equations with $c_0 = 0$ can be rewritten as $\{x, t\} = 2b_1(t)$, where $\{x,
t\}$ is the
{\it Schwarzian derivative} \cite{LG99} of the function $x(t)$ with respect to $t$.

Let us write KS-3 equations as a first-order system
\begin{equation}\label{firstKS3}
\left\{\begin{aligned}
\frac{dx}{dt}&=v,\\
\frac{dv}{dt}&=a,\\
\frac{da}{dt}&=\frac 32 \frac{a^2}v-2c_0(x)v^3+2b_1(t)v,
\end{aligned}\right.
\end{equation}
in the open submanifold $\mathcal{O}_2=\{(x,v,a)\in {\rm T}^2\mathbb{R}\mid
v\neq 0\}$ of ${\rm T}^2\mathbb{R}\simeq \mathbb{R}^3$, the referred to as {\it
second-order tangent bundle} \cite{MFM09} of $\mathbb{R}$.
 
Consider now the set of vector fields on $\mathcal{O}_2$ given by
\begin{equation}\label{VFKS1}
\begin{aligned}
N_1&=2v\frac{\partial}{\partial a},\\
N_2&=v\frac{\partial}{\partial v}+2a\frac{\partial}{\partial a},\\
N_3&=v\frac{\partial}{\partial x}+a\frac{\partial}{\partial v}+\left(\frac 32
\frac{a^2}v-2c_0(x)v^3\right)\frac{\partial}{\partial a},
\end{aligned}
\end{equation}
 which satisfy
the commutation relations
\begin{equation}
[N_1,N_3]=2N_2,\quad [N_1,N_2]=N_1,\quad [N_2,N_3]=N_3.
\end{equation}
Thus, they span a three-dimensional Lie algebra of vector fields $V$ isomorphic
to $\mathfrak{sl}(2,\mathbb{R})$. Since (\ref{firstKS3})  is determined by the
$t$-dependent vector field
$$
X_t=v\frac{\partial}{\partial x}+a\frac{\partial}{\partial v}+\left(\frac 32
\frac{a^2}v-2c_0(x)v^3+2b_1(t)v\right)\frac{\partial}{\partial
a},
$$
we can write $X_t=N_3+b_1(t)N_1.$ Consequently, $X$ takes values in the 
finite-dimensional Vessiot--Guldberg Lie algebra $V$ and becomes an $\mathfrak{sl}(2,\mathbb{R})$-Lie system. 
This generalizes the result provided in
\cite{GL12} for $c_0(x)=const.$

We shall now reduce the integration of (\ref{firstKS3}) with $c_0(x)=const.$, and in consequence the integration of the related (\ref{KS3}),
to working out a
particular solution of the Lie system (\ref{reduced}). To do so, we employ the Lie group action
$\varphi_{3KS}:SL(2,\mathbb{R})\times \mathcal{O}_2\rightarrow  \mathcal{O}_2$
whose
infinitesimal action is given by the Lie algebra isomorphism
$\rho:\mathfrak{sl}(2,\mathbb{R})\rightarrow V$ satisfying that
$\rho({\rm a}_\alpha)=N_\alpha$, with $\alpha=1,2,3$. This Lie group action holds that 
$$
\frac{d}{ds}\varphi_{3KS}(\exp(-s{\rm a}_\alpha),{\bf
t}^2_x)=N_\alpha(\varphi_{3KS}(\exp (-s{\rm a}_\alpha),{\bf t}^2_x)),
$$
with ${\bf t}^2_x\equiv (x,v,a) \in\mathcal{O}_2$ and $\alpha=1,2,3$.
Integrating $N_1$ and $N_2$,
we easily see that 
$$
\varphi_{3KS}\left(\exp(-\lambda_1{\rm a}_1),{\bf
t}_{x}^2\right)=\left(\begin{array}{c}
x\\
v\\
a+2v\lambda_1\\ 
\end{array}\right)$$
and
$$
\varphi_{3KS}\left(\exp(-\lambda_2{\rm a}_2),{\bf t}_{x}^2\right)=
\left(\begin{array}{c}
x\\
ve^{\lambda_2}\\
ae^{2\lambda_2}\\ 
\end{array}\right).
$$
To integrate $N_3$, we need to obtain the solutions of
\begin{equation}\label{Sys3}
\frac{dx}{d\lambda_3}=v,\qquad \frac{dv}{d\lambda_3}=a,\qquad \frac{da}{d\lambda_3}=\frac 32\frac {a^2}v-2c_0v^3.
\end{equation}
Proceeding, {\it mutatis mutandis}, as in the analysis of system (\ref{LC}), we obtain
$$
v(\lambda_3)\!=
\!\frac{v(0)}{F_{\lambda_3}(x(0),v(0),a(0))},
$$
with 
$$
F_{\lambda_3}({\bf
t}^2_{x})=\left(1-\frac{a\lambda_3}{2v}\right)^2+c_0v^2\lambda_3^2.
$$
Taking into account this and the first two equations within (\ref{Sys3}), we see that 
\begin{equation*}
\varphi_{3KS}\left(e^{-\lambda_3{\rm a}_3},
{\bf t}_{x}^2\right)=\left(
\begin{array}{c}
x+v\int^{\lambda_3}_0F^{-1}_{\lambda'_3}({\bf t}_{x}^2)d\lambda'_3\\
F^{-1}_{\lambda_3}({\bf t}_{x}^2)v\\
v\partial (F^{-1}_{\lambda_3}({\bf t}_{x}^2))/\partial \lambda_3
\end{array}\right).
\end{equation*}
Using decomposition (\ref{decomposition}), we can reconstruct the new action 
$$\varphi_{3KS}\left(g,
{\bf t}_{x}^2\right)=\left(
\begin{array}{c}
x+v\int^{\gamma/\alpha}_0\bar F^{-1}_{\lambda_3,g}({\bf
t}^2_x)d\lambda_3\\
\bar{F}^{-1}_{\gamma/\alpha,g}({\bf t}^2_x)v\\
v\frac{\partial (\bar{F}^{-1}_{\lambda_3,g}({\bf t}^2_x))}{\partial
\lambda_3}\big|_{\lambda_3=\gamma/\alpha}
\end{array}\right),
$$
with $\bar F_{\lambda_3,g}({\bf t}_x^2)=\alpha^{-2}F_{\lambda_3}(x,v\alpha^2,(a\alpha-2v\beta)\alpha^3\lambda_3)$, i.e., 
$$
\bar{F}_{\lambda_3,g}({\bf
t}^2_{x})=\left(\frac{1}{\alpha}-\frac{a\alpha -2v\beta}{2 v}\lambda_3\right)^2
+c_0v^2\alpha^2\lambda_3^2.
$$
This action enables us to write the general solution of
(\ref{firstKS3}) as
$$(x(t),v(t),a(t))=\varphi_{3KS}(g(t),{\bf t}^2_{x}),$$where ${\bf t}^2_{x}\in
\mathcal{O}_2$ and $g(t)$ is a particular solution of the equation on 
$SL(2,\mathbb{R})$ given by (\ref{reduced}). Hence, if ${\tau^{2)}}:(x,v,a)\in
{\rm T}^2\mathbb{R}\mapsto x\in
\mathbb{R}$ is the fiber bundle projection corresponding to the second-order
tangent
bundle on $\mathbb{R}$, we can write the general solution of
(\ref{KS3}) in the form
$$
x(t)=\tau^{2)}\circ \varphi_{3KS}(g(t),{\bf t}^2_{x}),
$$
where $g(t)$ is any particular solution of (\ref{reduced}).

Conversely, given the general solution of (\ref{KS3}), we can obtain a particular solution of (\ref{reduced}). As before, we focus on obtaining the particular solution $g_1(t),$ with $g_1(0)=e$. In this case, 
given two particular solution $x_1(t),x_2(t)$ of (\ref{KS3}) with $d^2 x_1/dt^2(0)=d^2x_2/dt^2(t)=0$, we obtain that the $t$-dependent coefficients $\alpha(t)$, $\beta(t)$, $\gamma(t)$ 
and $\delta(t)$ corresponding to the matrix expression of $g_1(t)$ obey a system similar to (\ref{sys1}) where $v$ and $x$ have been replaced by $a$ and $v$, respectively. 

\section{On the relations of Kummer--Schwarz equations with other equations}
We have already shown that the general solution of second- and third-order
Kummer--Schwarz equations can be obtained from a particular solution of a Lie
system in $SL(2,\mathbb{R})$ and vice versa. In this section, we will show that this property
is shared by all systems that are known to be closely related to
KS-2 and KS-3 equations: e.g., time-dependent frequency
harmonic
oscillators, Milne--Pinney and Riccati equations \cite{GGG11,BR97,Be07,Co94,SIGMA}. This allows
us to explain why the integration of one of these systems amounts to integrating
a particular instance of all the others. In addition, we find new remarkable systems of differential
equations that are related in this same way to second- and third-order Kummer--Schwarz
equations.

Let us start by analyzing the Riccati equations of the form
\begin{equation}\label{Ricc2}
\frac{dx}{dt}=b_1(t)+x^2,
\end{equation}
These equations are determined by a time-dependent vector field
$$
W_t=(b_1(t)+x^2)\frac{\partial}{\partial x},
$$
which can be written as $X_t=W_3+b_1(t)W_1$, where
$$
W_1=\frac{\partial}{\partial x},\qquad W_2=x\frac{\partial}{\partial x},\qquad
W_3=x^2\frac{\partial}{\partial x}
$$
satisfy the commutating relations
\begin{equation}\label{WRel}
[W_1,W_3]=2W_2,\quad [W_1,W_2]=W_1,\quad [W_2,W_3]=W_3.
\end{equation}
Hence, equations (\ref{Ricc2}) are Lie systems related to a Vessiot--Guldberg Lie Lie algebra isomorphic to
$\mathfrak{sl}(2,\mathbb{R})$. This Lie algebra gives rise to a local Lie group action
$\varphi:SL(2,\mathbb{R})\times\mathbb{R}\rightarrow\mathbb{R}$ of the form
$$
\varphi\left(\left(\begin{array}{cc}
\alpha&\beta\\\gamma&\delta\end{array}\right),x\right)=\frac{\alpha
x-\beta}{-\gamma x+\delta},\qquad
\alpha\delta-\beta\gamma=1,
$$
whose fundamental vector fields associated with ${\rm a}_1,{\rm a}_2$ and ${\rm a}_3$ are $W_1,W_2$ and $W_3$, respectively. 
In consequence, the solution of $W$ can be put in the form $x(t)=\varphi(g(t),x_0)$, with
$x_0\in\mathbb{R}$ and $g(t)$ given by (\ref{reduced}). Conversely, given three different particular solutions $x_1(t)$, $x_2(t)$ and $x_3(t)$ of (\ref{Ricc2}), 
we can easily determine $g(t)$, with $g(0)=e$, from the equations $x_i(t)=\varphi(g(t),x_i(0))$, with $i=1,2,3$.

Consider now the Milne--Pinney equations 
$$
\frac{d^2x}{dt^2}=-b_1(t)x+\frac{c}{x^3},
$$
with $c\in\mathbb{R}$ and $b_1(t)$ being an arbitrary $t$-dependent function. It is remarkable that
when $c=0$, we have a $t$-dependent frequency harmonic oscillator.
The Milne--Pinney equations in form of a first-order
system
\begin{equation}\label{osc}
\left\{\begin{aligned}
\frac{dx}{dt}&=v,\\
\frac{dv}{dt}&=-b_1(t)x+\frac{c}{x^3}
\end{aligned}\right.
\end{equation}
is governed by the time-dependent vector field
$W=W_3+b_1(t)W_1$, where
\begin{equation}\label{linear}
\begin{gathered}
W_1\!=\!-x\frac{\partial}{\partial v},\qquad W_2\!=\!\frac
12\left(v\frac{\partial}{\partial v}-x\frac{\partial}{\partial x}\right),\qquad
W_3\!=\!v\frac{\partial}{\partial x}+\frac{c}{x^3}\frac{\partial}{\partial v}.
\end{gathered}
\end{equation}
For simplicity, we restrict ourselves to the case $x>0$ and $c>0$. The above vector fields close on a Lie algebra isomorphic to
$\mathfrak{sl}(2,\mathbb{R})$ and give rise to the Lie group action $\varphi_{MP}:(A,(x,v))\in SL(2,\mathbb{R})\times {\rm T}\mathbb{R}_+\mapsto (\bar x,\bar v)\in {\rm T}\mathbb{R}_+$ given by
\begin{gather*}
\bar x =\sqrt{\dfrac{c+\left[(\alpha v+\beta x)(\gamma
      v+\delta x)+ c({\alpha\gamma}/{x^2})\right]^2}{(\alpha
    v+\beta x)^2+c{\alpha^2}/{x}^2}},\\
\bar v =\kappa \sqrt{\left(\alpha v+\beta
  x\right)^2+\dfrac{c\alpha^2}{x^2}\left(1-\dfrac{x^2}{\alpha^2\bar
    x^2}\right)},
\end{gather*}
where 
$$\kappa={\rm sign}\left(\frac{c\alpha\gamma}{x^2}+(\alpha v+\beta x)(\gamma v+\delta x)\right).
$$ 
This expression can be obtained proceeding as in previous sections or as in \cite{SIGMA}. This shows that
the general solution to (\ref{osc}) can be obtained through a particular solution to (\ref{reduced}). Conversely,
given two particular solutions $(x_1(t),v_1(t))$ and $(x_2(t),v_2(t))$ to (\ref{osc}), e.g. those with $v_1(0)=v_2(0)=0$, a particular solution to (\ref{reduced}) with $g(0)=e$ can be obtained by solving the algebraic system of equations $\varphi_{MP}(g(t),(x_i(0),v_i(0)))=(x_i(t),v_i(t))$, where $i=1,2$. 

Let us analyze a last example of Lie system related to
$\mathfrak{sl}(2,\mathbb{R})$. Consider 
\begin{equation}\label{ControlSys}\left\{
\begin{aligned}
\frac{dx}{dt}&=b_1(t)+x^2,\\
\frac{dy}{dt}&=2x,\\
\frac{dz}{dt}&=-e^y.
\end{aligned}\right.
\end{equation}
appearing in the application of the Wei-Norman method to $\mathfrak{sl}(2,\mathbb{R})$-Lie systems 
\cite{Dissertationes,Pi12}. We define
$$
W_1=\frac{\partial}{\partial x},\quad \!\!W_2=x\frac{\partial}{\partial
x}+\frac{\partial}{\partial y},\!\quad \! W_3=x^2\frac{\partial}{\partial
x}+2x\frac{\partial}{\partial y}-e^y\frac{\partial}{\partial z},
$$
which close on the commutation relations (\ref{WRel}). In view of these vector
fields, we easily see that system (\ref{ControlSys}) is a Lie system governed by
a time-dependent vector field $W_t=W_3+b_1(t)W_1$.

The integration of $W_1,W_2$ and $W_3$ results in an action
$\varphi_C:SL(2,\mathbb{R})\times \mathbb{R}^3\rightarrow \mathbb{R}^3$ of the
form
$$
\!\varphi_C\!\left(\!\left(\begin{array}{cc}
\alpha&\beta\\\gamma&\delta\end{array}\!\right),
\left(\!\begin{array}{c}
 x\\
y\\
z\\
\end{array}\right)\right)\!=\!\left(
\begin{array}{c}\frac{\alpha
x-\beta}{-\gamma x+\delta}\\y-\log\big (\delta-\gamma x)^2	\\\!
z-\frac{e^{y}\gamma}{\delta-\gamma x} \!\\\end{array}\right)\!,
$$
which allows us to write the general solution of (\ref{ControlSys}) as
\begin{equation}\label{sys4}
(x(t),y(t),z(t))=\varphi_C(g(t),(x_0,y_0,z_0)),
\end{equation}
with $g(t)$ being a particular solution of (\ref{eqLie}) and
$(x_0,y_0,z_0)\in\mathbb{R}^3$. Again, it is easy to prove that certain particular solutions to (\ref{ControlSys})
give rise to a solution of (\ref{reduced}) by solving the corresponding system induced by (\ref{sys4}).

Note that our results show that solving second- and third-order Kummer--Schwarz equations,
time-dependent frequency harmonic oscillators, Milne--Pinney equations, and
Riccati equations amounts to obtaining a particular solution of (\ref{reduced}). This 
provides a new geometric and unified explanation of the relations among the
solutions of these systems presented in different forms, e.g., through specific non-local changes of variables, in the literature \cite{AL08,Co94}. 
Moreover, our results can be potentially be extended to any $\mathfrak{sl}(2,\mathbb{R})$-Lie system whose Vessiot--Guldberg Lie algebra can be integrated into an action \cite{Dissertationes,RR96}.

\section{On the properties of the Schwarzian derivative}

The Schwarzian derivative of a real function $f=f(t)$ is
defined by
$$
\{f,t\}=\frac{d^3f}{dt^3}\left(\frac{df}{dt}\right)^{-1}-\frac
32\left[\frac{d^2f}{dt^2}\left(\frac{df}{dt}\right)^{-1}\right]^{2}.
$$
This derivative is clearly related to KS-3 equations (\ref{KS3}) with $c_0=0$, which can be written as $\{f,t\}=2b_1(t)$. 

Although a superposition rule for studying KS-3 equations was developed in \cite{CGL11}, the result provided in there was not
valid
when $c_0=0$, which retrieves the relevant equation $\{x,t\}=2b_1(t)$. This is why we aim to reconsider this case and its important connection to the Schwarzian derivative.

We shall now follow the method detailed in the introduction to obtain a superposition rule for (\ref{firstKS3}). The vector fields $N_1,
N_2,N_3$ are linearly independent at a generic point of
$\mathcal{O}_2\subset{\rm T}^2\mathbb{R}_0$.
Therefore, obtaining a superposition rule for (\ref{firstKS3}) amounts to
obtaining three functionally independent first-integrals common to all diagonal
prolongations $\widetilde N_1,\widetilde N_2,\widetilde N_3$ in
$(\mathcal{O}_2)^2$ satisfying (\ref{cond}). As $[\widetilde N_1,\widetilde N_3]=2\widetilde N_2$, it
suffices to obtain common first-integrals for $\widetilde N_1,\widetilde
N_3$ to describe first-integrals common to the integrable distribution $\mathcal{D}$
spanned by $\widetilde N_1,\widetilde N_2,\widetilde N_3$.

Let us start by solving $\widetilde N_1F=0$, with
$F:\mathcal{O}_2\rightarrow\mathbb{R}$, i.e.,
$$
v_0\frac{\partial F}{\partial a_0}+v_1\frac{\partial F}{\partial a_1}=0.
$$
The method of characteristics shows that $F$ must be constant along the
solutions of the associated {\it Lagrange--Charpit equations} \cite{De97}, namely
$$
\frac{da_0}{v_0}=\frac{da_1}{v_1},\qquad dx_0=dx_1=dv_0=dv_1=0.
$$
Such solutions are the curves $(x_0(\lambda),v_0(\lambda),a_0(\lambda),x_1(\lambda),v_1(\lambda),a_1(\lambda))$ within $\mathcal{O}_2$ with $\Delta=v_1(\lambda)a_0(\lambda)-a_1(\lambda)v_0(\lambda)$, for a real constant $\Delta\in\mathbb{R},$ and constant
$x_i(\lambda)$ and $v_i(\lambda)$, with $i=0,1$. In other words, there exists a function
$F_2:\mathbb{R}^5\rightarrow\mathbb{R}$ such that
$F(x_0,v_0,a_0,x_1,v_1,a_1)=F_2(\Delta,x_0,x_1,v_0,v_1)$.

If we now impose $\widetilde N_3F=0$, we obtain
$$
\widetilde N_3F=\widetilde N_3F_2=\frac{\Delta+a_1v_0}{v_1}\frac{\partial
F_2}{\partial v_0}+a_1\frac{\partial F_2}{\partial v_1}+v_0\frac{\partial
F_2}{\partial x_0}+v_1\frac{\partial F_2}{\partial x_1}+\frac{3\Delta^2+6\Delta
a_1v_0}{2v_1v_0}\frac{\partial F_2}{\partial\Delta}=0.
$$
We can then write that $\widetilde N_2F_2=(a_1/v_1)\Xi_1F_2+\Xi_2F_2=0$, where
$$
\Xi_1=v_0\frac{\partial}{\partial v_0}+v_1\frac{\partial}{\partial
v_1}+3\Delta\frac{\partial}{
\partial \Delta},\qquad \Xi_2=v_0\frac{\partial}{\partial x_0}+v_1\frac{\partial}{\partial
x_1}+\frac{\Delta}{v_1}\frac{\partial}{\partial
v_0}+\frac{3\Delta^2}{2v_0v_1}\frac{\partial}{\partial \Delta}.
$$
As $F_2$ does not depend on $a_1$ in the chosen coordinate system, it follows
$\Xi_1F_2=\Xi_2F_2=0$. Using the characteristics method again, we obtain that
$\Xi_1F_2=0$ implies the existence of a new function
$F_3:\mathbb{R}^4\rightarrow\mathbb{R}$ such that
$F_2(\Delta,x_0,x_1,v_0,v_1)=F_3(K_1\equiv v_1/v_0,K_2\equiv v_0^3/\Delta,
x_0,x_1)$. 

The only condition remaining is $\Xi_2F_3=0$. In the local coordinate system
$\{K_1,K_2,x_0,x_1\}$, this equation reads
$$
v_0\left(\frac 3{2K_1}\frac{\partial F_3}{\partial
K_2}-\frac{1}{K_2}\frac{\partial F_3}{\partial K_1}+\frac{\partial F_3}{\partial
x_0}+K_1\frac{\partial F_3}{\partial x_1}\right)=0,
$$  
and its Lagrange--Charpit equations becomes
$$
-K_2dK_1=\frac{2K_1dK_2}{3}={dx_0}=\frac{dx_1}{K_1}.
$$
From the first equality, we obtain that $K_1^3K_2^2=\Upsilon_1$ for a
certain real constant $\Upsilon_1$. In view of this and with the aid of the
above system, it turns out
$$
\frac 23 K_1^2dK_2=dx_1\longrightarrow \frac
23\Upsilon_1^{2/3}K_2^{-4/3}dK_2=dx_1.
$$
Integrating, we see that $-2K_2K_1^2-x_1=\Upsilon_2$ for a certain real
constant
$\Upsilon_2$. Finally, these previous results are used to solve the last part of the
Lagrange--Charpit system, i.e.,
$$
dx_0=\frac{dx_1}{K_1}=\frac{4\Upsilon_1dx_1}{(x_1+\Upsilon_2)^2}\longrightarrow
\Upsilon_3=x_0+\frac{4\Upsilon_1}{x_1+\Upsilon_2}.
$$
Note that $\partial(\Upsilon_1,\Upsilon_2,\Upsilon_3)/\partial(x_0,v_0,a_0)\neq 0$. Therefore,
considering $\Upsilon_1=k_1$, $\Upsilon_2=k_2$ and $\Upsilon_3=k_3$, we can 
obtain a mixed superposition rule. From these equations, we easily obtain 
\begin{equation}\label{sup0}
x_0=\frac{x_1k_3+k_2k_3-4k_1}{x_1+k_2}. 
\end{equation}
Multiplying numerator and denominator of the right-hand side by a non-null constant $\Upsilon_4$, the above expression 
can be rewritten as
\begin{equation}\label{sup1}
x_0=\frac{\alpha x_1+\beta}{\gamma x_1+\delta},
\end{equation}
with $\alpha=\Upsilon_4k_3,\beta=\Upsilon_4(k_2k_3-4k_1), \gamma=\Upsilon_4,\delta=k_2\Upsilon_4$. Observe that
$$
\alpha\delta-\gamma\beta=\Upsilon_4^2\Upsilon_1=\frac{\Upsilon_4^2v^3_0v_1^3}{(v_1a_0-a_1v_0)^2}\neq 0.
$$
Then, choosing an appropriate $\Upsilon_4$, we obtain that (\ref{sup0}) can be rewritten as (\ref{sup1}) for a family of constants $\alpha,\beta,\gamma,\delta$ such that $\alpha\delta-\gamma\beta=\pm 1$. It is important to recall that the matrices
$$
\left(\begin{array}{cc}
\alpha&\beta\\
\gamma&\delta\\
\end{array}\right),\qquad I=\alpha\delta-\beta\gamma=\pm 1,
$$
are the matrix description of the Lie group $PGL(2,\mathbb{R})$.

Operating, we also obtain that
$$
v_0=\frac{Iv_1}{(\gamma x_1+\delta)},\qquad a_0=I\left[\frac{a_1}{(\gamma x_1+\delta)^2}-\frac{2v_1^2\gamma}{(\gamma x_1+\delta)^3}\right].
$$
The above expression together with (\ref{sup1}) become a superposition rule for KS-3 equations with $c_0=0$ (written as a first-order system). In other words, the general solution $(x(t),v(t),a(t))$ of (\ref{KS3}) with $c_0=0$ can be written as
$$
(x(t),v(t),a(t))=\Phi(A,x_1(t),v_1(t),a_1(t)),
$$
with $(x_1(t),v_1(t),a_1(t))$ being a particular solution, $A\in PGL(2,\mathbb{R})$ and
$$
\Phi(A,x_1,v_1,a_1)=\left(\frac{\alpha x_1+\beta }{\gamma x_1+\delta},\frac{Iv_1}{(\gamma x_1+\delta)^2},I\left[\frac{a_1(\gamma x_1+\delta)-2v_1^2\gamma}{(\gamma x_1+\delta)^3}\right]\right).
$$
Moreover, $x(t)$, which is the general solution of a KS-3 equation with $c_0=0$, can be determined out of a particular solution $x_1(t)$ and three constants through
\begin{equation}\label{Basic}
x(t)=\tau^{2)}\circ\Phi\left(A,x_1(t),\frac{dx_1}{dt}(t),\frac{d^2x_1}{dt^2}(t)\right),
\end{equation}
where we see that the right-hand part does merely depend on $A$ and $x_1(t)$. This constitutes a {\it basic superposition rule} \cite{CGL11} for equations $\{x(t),t\}=2b_1(t)$, i.e., it is an expression that allows us to
describe the general solution of any of these equations in terms of a particular solution (without involving its derivatives) and some constants to be related to initial conditions. 
We shall now employ this superposition rule to describe some properties of the
Schwarzian derivative.

From the equation above, we analyze the relation between two particular solutions $x_1(t)$ and $x_2(t)$ of the same equation $\{x,t\}=2b_1(t)$, i.e., $\{x_1(t),t\}=\{x_2(t),t\}$. Our basic superposition rule (\ref{Basic}) tells us that from  $x_1(t)$ we can generate every other solution of the equation. In particular, there must exist certain real constants $c_1,c_2,c_3,c_4$ such that 
$$
x_2(t)=\frac{c_1 x_1(t)+c_2}{c_3 x_1(t)+c_4},\qquad c_1c_4-c_2c_3\neq 0.
$$ 
In this way, we recover a relevant property of this type of equations \cite{OT09}. 

Our basic superposition rule (\ref{Basic}) also provides us with information about the Lie symmetries of 
$\{x(t),t\}=2b_1(t)$. Indeed, note that (\ref{Basic}) implies that the local Lie group action
$\varphi:PGL(2,\mathbb{R})\times \mathbb{R}\rightarrow \mathbb{R}$ 
$$
\varphi(A,x)=\frac{\alpha x+\beta}{\gamma x+\delta},
$$
transforms solutions of $\{x(t),t\}=2b_1(t)$ into solutions of the same equation. The prolongation \cite{Ol93,Bo01} $\widehat\varphi: PGL(2,\mathbb{R})\times {\rm T}^2\mathbb{R}_0\rightarrow {\rm T}^2\mathbb{R}_0$ of $\varphi$ to ${\rm T}^2\mathbb{R}_0$, i.e.,
$$
\widehat\varphi(A,{\bf t}^2_x)\!=\!\left(\frac{\alpha x+\beta}{\gamma x+\delta},\frac{Iv}{(\gamma x+\delta)^2},I\frac{a(\gamma x+\delta)-2\gamma v^2}{(\gamma x+\delta)^3}\right),
$$
gives rise to a group of symmetries $\varphi(A,\cdot)$ of (\ref{firstKS3}) when $c_0=0$. The fundamental vector fields of this action are spanned
by
$$
\begin{gathered}
Z_1=-\frac{\partial}{\partial x},\qquad Z_2=x\frac{\partial}{\partial x}+v\frac{\partial}{\partial v}+a\frac{\partial}{\partial a},\qquad
Z_3=-\left(x^2\frac{\partial}{\partial x}+2vx\frac{\partial}{\partial v}+2(ax+v^2)\frac{\partial}{\partial a}\right),
\end{gathered}
$$
which close on a Lie algebra of vector fields isomorphic to $\mathfrak{sl}(2,\mathbb{R})$ and commute with $X_t$ for every $t\in\mathbb{R}$. In addition, their projections onto $\mathbb{R}$ must be Lie symmetries of $\{x(t),t\}=2b_1(t)$. Indeed, they read
$$
S_1=-\frac{\partial}{\partial x},\qquad S_2=x\frac{\partial}{\partial x},\qquad S_3=-x^2\frac{\partial}{\partial x},
$$
which are the known Lie symmetries for these equations \cite{OT05}.

Consider now the equation $\{x(t),t\}=0$. Obviously, this equation admits the
particular solution $x(t)=t$. This, together with our basic superposition rule, show
that the general solution of this equation is
\begin{equation}\label{symsol}
x(t)=\frac{\alpha t+\beta}{\gamma t+\delta},\qquad \alpha\delta-\gamma\beta\neq 0,
\end{equation}
recovering another relevant known solution of these equations. 

\section{The Kummer--Schwarz equations and mixed superposition rules}
In previous sections, we have provided an alternative and unified approach for the
study of relations between Kummer--Schwarz equations and other 
remarkable differential equations. In this section, we want to show that
when we describe all these equations by a Lie system
with a Vessiot--Guldberg Lie algebra isomorphic to
$\mathfrak{sl}(2,\mathbb{R}),$ we can retrieve and generalise certain
expressions found in the literature. With this aim, we shall now apply the fundamentals explained in Section
\ref{Fun} to
derive a mixed superposition rule for KS-2 equations. 

As both (\ref{VFKS1}) and (\ref{linear}) close on the same commutation
relations, we can build up the family of vector fields $\widehat
W_\alpha=W_\alpha\times W_\alpha$, for $\alpha=1,2,3$
\begin{equation}
\begin{aligned}
\widehat X_1&=-x_1\frac{\partial}{\partial v_1}-x_2\frac{\partial}{\partial
v_2},\\
\widehat X_2&=\frac 12\left(v_1\frac{\partial}{\partial
v_1}+v_2\frac{\partial}{\partial v_2}-x_1\frac{\partial}{\partial
x_1}-x_2\frac{\partial}{\partial x_2} \right),\\
\widehat X_3&=v_1\frac{\partial}{\partial x_1}+v_2\frac{\partial}{\partial x_2},
\end{aligned}
\end{equation}
that are linearly independent at a generic point of $\mathbb{R}^4$.
Consequently, the vector fields $\widetilde X_\alpha=\widehat
X_\alpha\times M_\alpha	$, with $\alpha=1,2,3$, namely
\begin{equation}
\begin{aligned}\label{1}
\!\widetilde X_1\!&=-x_1\frac{\partial}{\partial
v_1}-x_2\frac{\partial}{\partial v_2}+2x\frac{\partial}{\partial v},\\
\!\widetilde X_2\!&=\!x\frac{\partial}{\partial
x}+2v\frac{\partial}{\partial
v}\!+\!\frac 12\!\left(\!v_1\frac{\partial}{\partial
v_1}\!+\!v_2\frac{\partial}{\partial v_2}\!-\!x_1\frac{\partial}{\partial
x_1}\!-\!x_2\frac{\partial}{\partial x_2} \right),\\
\!\widetilde X_3\!&=v_1\frac{\partial}{\partial x_1}+v_2\frac{\partial}{\partial x_2}+v\frac{\partial}{\partial x}+\left(\frac 32
\frac{v^2}{x}-2c_0x^3\right)\frac{\partial}{\partial v},
\end{aligned}
\end{equation}
are also linearly independent at a generic point of 
$\mathbb{R}^6$. Besides, a mixed superposition rule for
(\ref{FirstOrderKummer}) can be obtained by determining a family of two
functionally independent functions $F_1$ and $F_2$ playing the r\^ole of common
first-integrals of (\ref{1}) and such that $\partial (F_1,F_2)/\partial(x,v)\neq 0.$
As $[\widetilde X_1,\widetilde X_3]=2\widetilde X_2$, we only need to derive
 first-integrals common to $\widetilde X_1$ and $\widetilde
X_3$.  Let us assume that $F:\mathbb{R}^6\rightarrow \mathbb{R}$ is such a
first-integral. Then, the equation $\widetilde X_1F=0$ reads
$$
2x\frac{\partial F}{\partial v}-x_1\frac{\partial F}{\partial
v_1}-x_2\frac{\partial F}{\partial v_2}=0.
$$
Their corresponding Lagrange--Charpit equations show that $F$ must be constant
along the solutions of the system
$$
dx=dx_1=dx_2=0,\qquad \frac{dv}{2x}=\frac{dv_1}{-x_1}=\frac{dv_2}{-x_2}.
$$
In other words, $F$ is constant along curves with constant $x,x_1,x_2$ and such that
$x_1v+2xv_1=\xi_1$ and $x_2v+2xv_2=\xi_2$, 
for constants $\xi_1, \xi_2\in\mathbb{R}$. Consequently, $F(x,x_1,x_2,v,v_1,v_2)=F_2(x,x_1,x_2,\xi_1,\xi_2)$ for a function
$F_2:\mathbb{R}^5\rightarrow\mathbb{R}$. Using the coordinate system
$\{x,x_1,x_2,\xi_1,\xi_2,v\}$, we obtain that
$$
\widetilde X_3F=\widetilde X_3F_2=v\left[\frac{\partial F_2}{\partial
x}+\frac{3\xi_1}{2x}\frac{\partial F_2}{\partial
\xi_1}+\frac{3\xi_2}{2x}\frac{\partial F_2}{\partial
\xi_2}\right]+\sum_{i=1,2}\left(\frac{\xi_i-x_iv}{2x}\frac{\partial
F_2}{\partial
x_i}-2c_0x_ix^3\frac{\partial F_2}{\partial
\xi_i}\right)=0.
$$
The previous expression has to be satisfied for all $v$, and $F_2$ does not
depend on it. Taking into account that $\widetilde X_3F=\Theta_0F+v\Theta_1F=0$,  where
\begin{eqnarray*}
\Theta_0=\frac{\partial}{\partial x}+\frac{3\xi_1}{2x}\frac{\partial
F_2}{\partial \xi_1}-\frac{x_1}{2x}\frac{\partial }{\partial
x_1}+\frac{3\xi_2}{2x}\frac{\partial }{\partial
\xi_2}-\frac{x_2}{2x}\frac{\partial  }{\partial x_2},\\
\Theta_1=\frac{\xi_1}{2x}\frac{\partial }{\partial
x_1}+\frac{\xi_2}{2x}\frac{\partial  }{\partial x_2}-2c_0x_1x^3\frac{\partial
 }{\partial \xi_1}-2c_0x_2x^3\frac{\partial }{\partial \xi_2},
\end{eqnarray*}
we obtain
that $\Theta_0F_2=\Theta_1F_2=0$. From the first equation, we see that
$$
\frac{dx}{2x}=\frac{d\xi_1}{3\xi_1}=\frac{d\xi_2}{3\xi_2}=\frac{dx_1}{-x_1}
=\frac{dx_2}{-x_2}.
$$
Proceeding as above, we obtain that there must exist a function
$F_3:\mathbb{R}^4\rightarrow\mathbb{R}$ such that
$F_2(x,x_1,x_2,\xi_1,\xi_2)=F_3(\Gamma_1\equiv xx_1^2,\Gamma_2\equiv x
x_2^2,\Xi_1\equiv \xi_1^2/x^3,\Xi_2\equiv \xi_2^2/x^3)$.

Finally, using the coordinate system
$\{\Gamma_1,\Gamma_2,\Xi_1,\Xi_2\}$, we obtain that 
$\Theta_1F_2=\Theta_1F_3=0$ reads
$$
\sum_{i=1,2}\left(-4c_0\sqrt{\Xi_i\Gamma_i}\frac{\partial}{\partial
\Xi_i}+\sqrt{\Xi_i\Gamma_i}\frac{\partial}{\partial
\Gamma_i}\right)=0,
$$
which gives us two common first-integrals to $\widetilde X_1$ and $\widetilde
X_3$, i.e.,
$$
F_i\equiv\Xi_i+4c_0\Gamma_i=\frac{(x_iv+2xv_i)^2}{x^3}+4c_0xx_i^2,\quad i=1,2.
$$
Note that $\partial(F_1,F_2)/\partial(x,v)\neq 0$. Then, we can derive a superposition rule from the equations $F_1=I_1$ and $F_2=I_2$ for $I_1,I_2\in\mathbb{R}$. 
Using $F_1=I_1$, we obtain
\begin{equation}\label{v}
v=x\frac{\pm\sqrt{(I_1-4c_0xx_1^2)x}-2v_1}{x_1},
\end{equation}
and from $F_2=I_2$, we reach to
\begin{equation}\label{xSup}
x=4W^2\left[I_2x_1^2+I_1x_2^2\pm 2x_1x_2\sqrt{I_1I_2-4^2c_0W^2}\right]^{-1},
\end{equation}
where $W=x_1v_2-v_1x_2$ and $I_1I_2-4^2c_0W^2>0$. Plugging the above expression into (\ref{v}), we see that
\begin{equation}\label{ySup}
\!\!v\!=\!-8W^2\!\frac{I_2x_1v_1\!+\!I_1x_2v_2\!\pm\!
(x_1v_2\!+\!v_1x_2)[I_1I_2\!-\!4^2c_0W^2]^{\frac 12}}{\left[I_2x_1^2+I_1x_2^2\pm
2x_1x_2\sqrt{I_1I_2-4^2c_0W^2}\right]^{2}}.
\end{equation}
These two expressions permit us to write the general solution $(x(t),v(t))$ of (\ref{FirstOrderKummer}), i.e. a KS-2 equation written as a first-order system, in terms of two generic solutions $(x_1(t),v_1(t))$ and $(x_2(t),v_2(t))$ of (\ref{osc}), i.e.,
 a time-dependent frequency harmonic oscillator in first-order form, as 
$$
(x(t),v(t))=\Phi(x_1(t),v_1(t),x_2(t),v_2(t),I_1,I_2),
$$
where the components of $\Phi=(\Phi_x,\Phi_v)$ are given by (\ref{xSup}) and (\ref{ySup}), respectively.
 
Note also that, given two solutions $(x_1(t),v_1(t))$ and $(x_2(t),v_2(t))$
of (\ref{osc}), the expression $W=x_1(t)v_2(t)-v_1(t)x_2(t)$ is a constant of motion.
Then, we can redefine the constants as $k_2=I_1/4W^2$ and $k_1=I_2/4W^2$ to obtain an equivalent superposition rule
$$
\begin{gathered}
x=\left[k_1x_1^2\!+\!k_2x_2^2\pm 2x_1x_2\sqrt{k_1k_2-c_0W^{-2}}\right]^{-1},\\
v=\!-2\!\frac{k_1x_1v_1\!+\!k_2x_2v_2\!\pm\!
(x_1v_2+v_1x_2)\sqrt{k_1k_2\!-c_0W^{-2}}}{\left[k_1x_1^2+K_2x_2^2\pm
2x_1x_2\sqrt{k_1k_2-c_0W^{-2}}\right]^{2}}.\\
\end{gathered}
$$

It is straightforward that the first expression above allows us to describe the general solution of KS-2 equations by means of two generic solutions of $t$-dependent
frequency harmonic oscillators and their first-order derivatives. 
This result generalizes the expression obtained by Berkovich \cite{Be07} for equations with
$b_1(t)=const.$, which is only valid for two particular solutions of
the harmonic oscillator with Wronskian equal to one.

\section{Conclusions and Outlook}

We have provided a new geometric approach to Kummer--Schwarz equations that
retrieves previous results in a simpler and unified way. Among our future endeavors, we shall lay some attention upon the structure of
Lie symmetries and develop discrete methods for the study of Lie systems, 
which could be of great use in the study of Kummer-Schwarz equations
as well.

To conclude, we are very interested in developing the theory of {\it partial superposition
rules} \cite{CGM07}, as much as we are in the generalization of other expressions describing
general solutions of Kummer--Schwarz equations \cite{Be07} in terms of certain
specific sets of particular solutions.

\section*{Acknowledgments}
The research of the authors was
supported by the Polish National Science Centre
under the grant HARMONIA Nr 2012/04/M/ST1/00523.
Research of J. de Lucas has been partially financed by research projects FMI24/10
(DGA) and MTM2010-12116-E. C. Sard\'on acknowledges a fellowship provided by the
University of Salamanca and partial financial support by research project
FIS2009-07880
(DGICYT).



\end{document}